# In Pursuit of Privacy: The Value-Centered Privacy Assistant

**Sarah E. Carter**
TU Delft
The Netherlands
S.E.Carter@tudelft.nl
https://orcid.org/0000-0003-3621-5962

**Dayana Spagnuelo**
TNO
The Netherlands
dayana.spagnuelo@tno.nl
https://orcid.org/0000-0001-6882-6480

**Kathryn Cormican**
University of Galway
Ireland
kathryn.cormican@universityofgalway.ie
https://orcid.org/0000-0003-1688-1087

**Mathieu d'Aquin**
LORIA
Université de Lorraine
France
mathieu.daquin@loria.fr
https://orcid.org/0000-0001-7276-4702

**Ilaria Tiddi**
Vrije Universiteit Amsterdam
The Netherlands
i.tiddi@vu.nl
https://orcid.org/0000-0001-7116-9338

**Heike Felzmann**
University of Galway
Ireland
heike.felzmann@universityofgalway.ie
https://orcid.org/0000-0002-7355-6451

## Abstract

Many users make quick decisions that affect their data privacy without due consideration of their values. One such decision is whether to download a smartphone app to their device. Previous work has suggested a relationship between values, privacy preferences, and app choices, and proposed a value-centered approach to privacy that conceptually unites these relationships. In this work, we translate this theory into practice by constructing a prototype smartphone value-centered privacy assistant (VcPA) – a privacy assistant system that promotes user privacy decisions based on personal values. To do this, we designed and conducted an online survey that captured values and privacy preferences when considering whether to download an app from 273 smartphone users. Using this data, we constructed VcPA user profiles by clustering survey data based on the value rankings and stated privacy preferences. We then tested the VcPA, using selective notices, a "suggest alternatives" feature, and exploratory notices, with 77 users in a synthetic Mock App Store (MAS) setting and conducted follow-up semi-structured interviews. We establish proof-of-concept that a VcPA helps users make more value-centered app choices and identified improvements so that an assistant can be deployed on smartphone app stores.

## Introduction

Many digital services, including app stores, are not designed to support users in making privacy decisions that are consistent with their values. While designing to promote value-centered decision-making can be viewed as critical for promoting user autonomy and agency over data privacy choices (Terpstra et al. 2019; Carter 2022 & 2023), data privacy notices and consent dialogs, in their current form, are not designed to promote such decisions. For example, certain privacy notices and policy design choices can prompt users to consent (Solove 2021), regardless of whether it is in line with their values to do so.

To help users make more meaningful privacy choices, a value-centered privacy assistant (VcPA) was suggested by Carter (2022 & 2023). The VcPA is based on an existing technology, personalized privacy assistants (PPAs) (Liu et al. 2016; Liu, Lin, and Sadeh 2014; Warberg, Acquisti, and Sicker 2019), which aim to assist users in making decisions more consistent with their privacy preferences. In contrast to privacy assistants, the VcPA bases privacy recommendations on user values rather than privacy preferences alone. This, in theory, should promote more autonomous decisions that are meaningfully aligned to a user's personal values.

In particular, the VcPA aims to address one critical privacy choice: whether to download a smartphone app. In addition to the rationale put forth in Carter (2022 & 2023) for choosing this privacy decision point for a value-centered intervention, additional literature pertaining to the relationship between values, smartphone, and/or privacy seems to suggest that the three are intertwined (Alashoor et al. 2015; Nurwidyantoro et al. 2022; Shams et al. 2023; Perera et al. 2019; Obie et al. 2021). For example, Nurwidyantoro et al. (2022) found that app value statements – for example, stressing the value of privacy for Signal and Focus – seemed to influence the values they identified when exploring relevant values for apps on GitHub discussion forms. In addition, possible links between values and the level of privacy concern were proposed by Alashoor et al. (2015). This suggests that company value statements can be influential when looking to choose value-consistent apps, and users' personal values, privacy preferences, and app choices are all interrelated in some manner. Secondly, smartphones and apps are modern-day utilities that permeate all aspects of life. Our increased dependence on smartphones is further fueled through an app's ability to "seduce" users into giving more data away through gamification and other strategies (Troullinou 2017), increasing their attractiveness when considering an exploration of privacy choices and values.

In this paper, we establish proof-of-concept that a VcPA system – constructed using values and privacy preferences – helps users make app choices in accordance with their personal values. To accomplish this goal, we firstly aimed to translate existing theoretical and conceptual VcPA work into a prototype system for choosing apps. We secondly aimed to evaluate the VcPA and its features by assessing usability, evaluating effectiveness, and identifying areas for improvement.

This works begins by presenting a theoretical lens for measuring and understanding values – the Theory of Basic Human Values (THBV) (Schwartz 1992)– necessary for grounding our exploration of values and privacy decisions. We then explore the relevant background literature around privacy decision-making, including the challenges faced by privacy notices and efforts to design for more meaningful, value-centered privacy decisions. After laying this groundwork, we discuss how we accomplished our aims of building a VcPA and gathering empirical data to establish VcPA usability and effectiveness. Firstly, we present how we adapted an established survey from the TBHV to include smartphone app privacy preferences. We then illustrate how we utilized the survey data to construct user profiles for a value-centered privacy assistant by clustering values and privacy preferences. Secondly, we detail how we constructed and tested a VcPA prototype in a synthetic online "Mock App Store" testing environment for privacy assistants, including results that establish the VcPA's usability and effectiveness. Lastly, we consider future research avenues to move the smartphone VcPA from proof-of-concept into app stores.

**Measuring Values: Theoretical Foundations**
According to the TBHV, values motivate us to act in pursuit of one of three universal human requirements as defined by evolutionary psychology: our individual needs as biological organisms; our coordinated social interactions; and to ensure the welfare of social groups (Schwartz 1992). Values refer to motivational goals and encourage us to act. Shalom Schwartz, who developed the theory, claims that there is a set of ten universal human values that are defined by their distinct motivational goal, organized in a quasi-circular manner in relation to each other (Schwartz 1992; 2012). These tensions and similarities can be grouped along two axes: 1.) Openness to Change (values comprise: *Stimulation*, *Self-Direction*, *Hedonism*) vs. Conservation (values comprise: *Security*, *Conformity*, *Tradition*), capturing the tension between our desire for individual independence and for order; and 2.) Self-Enhancement (values: *Power*, *Achievement*, *Hedonism*) vs. Self-Transcendence (values: *Universalism*, *Benevolence*), which captures the tension between our own interests and our concern for the welfare of others. Notably, *Hedonism* has aspects of both Openness to Change and Self-Enhancement (Schwartz 2012). These values further serve as standards by which we evaluate good and bad, and become activated (e.g., threatened or upheld) in an applicable context. Activation is further linked to our emotional responses. Everyone has a set of values whose relative priorities to each other remain the same, independent of the situation (Schwartz et al. 2012).

Established empirical measures using Schwartz values include the Schwartz Value Survey and the Short Schwartz Value Survey (SSVS) (Schwartz 1992; 1994; Lindeman and Verkasalo 2010), which have been used as a means of probing values in smartphone settings (Obie et al. 2021; Shams et al. 2023; Nurwidyantoro et al. 2022). For these measurements, the ten values are understood as a "continuum of motivations," not fully distinct but with reasonable predictive power for value analysis (Schwartz 1992).

**Background and Related Work**
Insights from behavioral and cognitive psychology suggest that privacy notices are not successful at fulling their purpose: eliciting informed consent. Many users "click through" privacy notices because they are overwhelmed by the amount they receive (Schaub et al. 2015). Besides this "notice fatigue," they can also fall prey to a host of cognitive biases — mental "shortcuts" used for fast decision-making (Solove 2021). Data collectors can encourage consent on privacy notices using "dark" or "deceptive" patterns on notice design (Gray et al. 2018; Mathur et al. 2019; Utz et al. 2019), such as cookie privacy notices that highlight "accept all" to nudge users to consent. A lack of relevant privacy controls also causes problems for users (Carter 2022; Terpstra et al. 2019; Carter 2023), where they are not always able to act in the manners they choose. Because notice fatigue, dark patterns, and a lack of relevant privacy control discourage mindful, value-centered privacy decisions, it is therefore doubtful that privacy notices elicit truly autonomous, informed consent.

Instead, some human-computer interaction research has emerged applying insights from cognitive and behavioral psychology to design for wellbeing (Cox et al. 2016; Peters, Calvo, and Ryan 2018; Sandhaus 2023; Terpstra et al.

2019). Particularly promising is utilizing friction – small obstacles during a user's interaction with technology – to promote wellbeing (Cox et al. 2016; Sandhaus 2023). In this case, researchers deploy friction to induce a shift from "fast" to "slow" user thinking and thereby facilitate more autonomous decision-making. As described by Kahneman (2011), "Fast" System 1 thinking is automatic, mindless, and ripe with heuristics and biases, while "Slow" System 2 thinking is conscious, deliberate, and mindful. By introducing this friction, researchers are aiming to promote more System 2 thinking and therefore more deliberate interactions with technology.

Similarly, Carter (2022, 2023) proposed designing for user privacy decisions by 1.) understanding them in a manner that brings the role of their personal values to the forefront, and 2.) utilizing this understanding to promote more meaningful privacy decisions. The author utilizes the Four-Dimensional Theory of Self-Governance (4DT) put forth by Suzy Killmister (2017) to conceptualize value-centered privacy decisions – in sum, exploring why users may not act according to their values and identifying areas where selective friction could encourage more value-centered privacy decisions.

Carter further proposes ideas for making a value-centered privacy assistant (VcPA). This assistant is inspired by personalized privacy assistants (PPAs) — smartphone or IoT machine learning-based assistants that utilize user privacy preference profiles to trigger privacy notices to help the user manage their privacy decisions (Das et al. 2018; Liu et al. 2016). However, a 4DT analysis of existing PPAs suggests that certain modifications would be required to better support autonomous, value-centered privacy choices (Carter, 2022; 2023) (Table 1). The VcPA improves upon PPAs by shifting the focus from user privacy preferences to users' personal values, aiming to facilitate data privacy decisions that align with one's values rather than their privacy preference alone. To start, the VcPA was conceptualized for one privacy decision point — when deciding whether to download a smartphone app from an app store. It consists of three features based on the 4DT-informed understanding of value-centered data privacy decisions and the aforementioned analysis of PPAs through a 4DT lens (Carter 2021; 2022; 2023). It utilizes *selective notices* based on a user's values, encompassed in a value profile that determines when notices are issued; encourages users to re-assess their profile choice using *exploratory notices*, which checks with the user that the selective notices they have been receiving are still relevant to them; and *suggests alternative applications* whose data collection practices are more consistent with their value profile (Table 1).

In this work, we expand on these conceptualizations of a VcPA to empirically evaluate a proof-of-concept VcPA prototype. In the following section, we detail how the VcPA was constructed and evaluated using a mixed-method approach: a value and preference survey based on the SVSS (Lindeman and Verkasalo 2010); user tests in a "Mock App Store" setting (Carter, Tiddi, and Spagnuelo 2022); and follow-up semi-structured interviews.

| Feature | Description |
|---|---|
| *Selective Notices* | Encouraging users to "slow down" before downloading an app by introducing notices that appear when an app's data collection practice is not consistent with a user's selected value profile |
| *Suggesting Alternative Applications* | A "suggestion page" of alternative applications with similar function that are more consistent with the user's value profile, linked to the selective notices |
| *Exploratory Notices* | Promoting users at periodic time intervals with a notice to check that their value profile and resulting selective notices are still relevant to them |

Table 1: Features of a VcPA, modified from Carter (2021).

## Methods

To construct and test a VcPA, a mixed-methods study was conducted. This was chosen to 1.) capture the relationship between values and privacy preferences to generate VcPA profiles; 2.) empirically measure the usability of the VcPA; and 3.) elicit user feedback. The study consisted of three phases, with consent and information sheets provided to participants before each phase.

*Phase 1.* To collect data on values, the SSVS survey (Lindeman and Verkasalo 2010) was adapted and verified to include questions about smartphone privacy preferences. This survey asked participants about their values and privacy preferences when considering whether to download an app.

*Phase II.* To test the VcPA, a testing environment, VcPA logics, and a study task were created. A testing environment, called the Mock App Store (MAS) (Carter, Tiddi, and Spagnuelo 2022), was further fined-tuned for VcPA testing to include an updated prototype VcPA system. Results from Phase I were used to craft value profiles for the VcPA. To test the VcPA system and elicit feedback, participants were asked to partake in an exercise on the MAS. The "Mock App Store Exercise" asked participants to browse the apps on the Store with the VcPA assisting them. They also provided their feedback on the experience with an exit survey.

*Phase III.* To complement the feedback obtained on the exit survey, we also conducted follow-up semi-structured interviews with a select number of MAS participants. This allowed us to further assess the VcPA and elicit feedback from participants not captured on the exit survey.

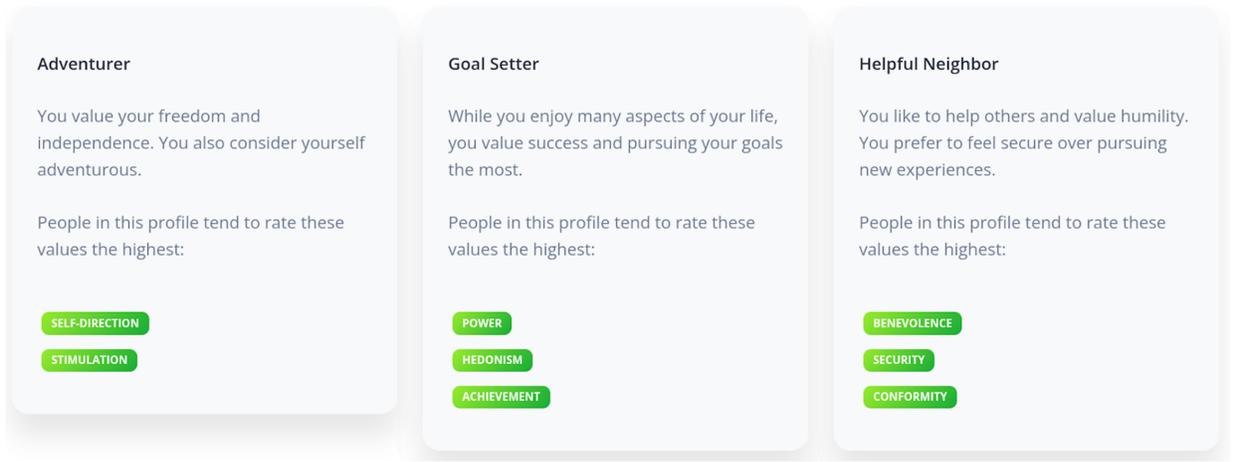

Figure 1: Presentation of VcPA profiles to participants on the Mock App Store (MAS).

**Phase I: Online Value and Privacy Preference Survey**
A survey was developed to quantify the relationship between user personal values, smartphone app choice, and privacy preferences for VcPA profile design. To measure values, we utilized an established method from the TBHV. To decrease the length of the online survey, the Schwartz Short Value Survey (SSVS) was modified in order to ask survey participants about their values (Lindeman and Verkasalo 2010). Based on the original 56/57 question Schwartz Value Survey (SVS) (Schwartz 1994; 1992), the SVSS asks participants directly about 10 broad values (e.g., "How important is *Hedonism* as a life guiding principle for you"?). Participants were asked to score values on a scale from 1 ("opposed") to 9 ("of supreme importance"), both in their lives more broadly and when considering whether to download an app. We created two versions of the survey with two example apps to give participants a tangible scenario of app choice. We used the health and fitness app Lose It! (https://www.loseit.com) for weight loss and the environmental app OpenLitterMap (https://openlittermap.com). Each participant was randomly assigned a version of the survey (Supplementary Table I).

To measure privacy preferences, we also included a series of questions based on Apple's Privacy Labels (https://www.apple.com/privacy/labels/), which categorize application data collection practices according to 1.) whether the data is linked (where linked means the data is linked to your identity) or tracked (data shared across companies or services, often for advertising purposes), and 2.) 13 types of data (health and fitness information, financial information, location, sensitive information, contacts, phone content, browsing history, search history, purchase history, usage data, general diagnostic data, contact information, other identifiers). The survey had one question for unlinked, linked, and tracked data, asking participants to select which types of data they would allow to be collected (Supplementary Table II). "Contact information" and "other identifiers" were only applicable to linked and tracking data. Collecting no data ("none") was also an option. This resulted in 38 different privacy preferences (e.g., unlinked location). Like the choice of the SSVS over the SVS, Apple (iOS) privacy permissions were utilized over Android for their ease of use. At the time the study was conducted, Android did not have a simplified, accessible list of their 300+ app privacy permissions (list available at: https://gist.github.com/Arinerron/1bcaadc7b1cbeae77de0263f4e15156f).

We then tested the survey. Survey feedback and testing aimed to encompass many disciplines due to the interdisciplinary nature of the study. Survey feedback and pre-testing was conducted with a group of colleagues from different disciplines at [home university]. Cumulatively, these individuals had a broad range of expertise, encompassing data science, law, engineering, human-computer interaction (HCI), quantitative and qualitative methodology, applied ethics, and philosophy. Based on survey pre-testing and expert feedback, the survey scale was modified from the original SVSS scale of -1 to 5 to a scale of 1 to 9 to add an intuitive midpoint at the number "5." We also included the definitions of the values first, followed by the value, to increase participant comprehension (Supplementary Table I). We also made a few small cosmetic alterations to the survey to increase attractiveness and understandability.

Participants were recruited using snowball sampling within the researchers' networks (e.g., via email or social media). Survey participants were required to be 18 years or older; native or fluent speakers of English; and currently own or previously owned a smartphone. The survey began with demographic questions related to age, gender, nationality, English proficiency, smartphone use, and education.

Data was pre-processed for consistency by ensuring that if a participant said they were willing to share a particular type of data for tracking, that the data reflected they also were willing to share linked and unlinked data of the same type. The same was done for a willingness to share linked data and unlinked. After the data was pre-processed, values and privacy preference correlations were analyzed using Spearman correlations.

### Phase II: Mock App Store Study

To conduct Phase II of the study, value profiles for the VcPA were designed, a testing environment for the prototype was established, and VcPA logics were embedded in the testing environment. To accomplish this, we first utilized data from Phase I to craft profiles. We then created a Mock App Store (MAS) by embedding updated VcPA logics and working profiles.

Firstly, value profiles for the VcPA were constructed using hierarchical clustering on survey data from the value and privacy preference survey. Hierarchical clustering has been used previously to identify privacy profiles for smartphone privacy assistants (Liu et al. 2016). Hierarchical clustering was conducted on app-specific values and stated general life-guiding principles. The app-specific clustering yielded two clusters of those who care about every value very highly and those who do not care much about any value. Hierarchical clustering based on general values (stated life-guiding principles) proved more promising. Following z-score normalization, used to account for sensitivity to magnitude (Milligan and Cooper 1988), hierarchical clustering based on general values yielded three clusters. Clustering was further verified by plotting profiles in three dimensions according to the variables (Schwartz values) with the highest variance (*Power, Hedonism*, and *Achievement*). The visual analysis of a 3D plot suggested that cluster 1 and cluster 2 tended to be distinguished from cluster 3 by higher *Hedonism* scores (Supplementary Figure 1). Clusters 1 and 2 were largely split along an *Achievement*/*Power* axis. Because Schwartz's theory of values postulates that the order in which we prioritize our values should be conserved across contexts ("trans-situational") (Schwartz 2012; de Wet, Wetzelhütter, and Bacher 2019), we selected these clusters for further profile development. Three profiles were designed from the clusters by crafting personas (Rosson and Carroll 2002) – in this case, narrative descriptions of each cluster based on their distinctive value characteristics (Figure 1, previous page). To guide us during persona creation, we used the Kushal Wallis test with a Dunn post-hoc test to identify general values that were ranked significantly higher or lower between clusters (p<0.05) (Kruskal and Wallis 1952). We also considered the values that were ranked highest within each profile. *Profile 1, Adventurer,* has significantly lower values scores for *Security*, *Tradition*, *Benevolence*, and *Conformity* compared to the other two clusters.

Its highest ranked values are *Stimulation* and *Self-direction*. *Profile 2, Goal Setter,* has significantly higher *Power*, *Achievement*, and *Hedonism* scores than the other two clusters. These are also its highest ranked values. *Profile 3, Helpful Neighbor,* has lower *Self-Direction*, *Stimulation*, and *Hedonism* compared to the other two clusters. Its highest-ranking values are *Benevolence, Security*, and *Conformity*.

Next, we gathered apps for the MAS. The MAS consisted of Apple App Store apps from one of the domains explored in the online survey, the "health and fitness" Apple app category. App metadata, keywords, data collection specifications, and similar applications were collected from the AppTweak API (https://www.apptweak.io). The top ten apps in this category were taken from the US Apple App Store on September 21st, 2021, and their app ID used to request similar apps in AppTweak. This group of similar apps became one "app family," and this process was conducted to have ten complete app families (for a total of 100 apps). We then requested all the metadata and keywords for the gathered apps. To deal with duplicates, we merged families that shared apps and used the Jaccard index to sort any remaining duplicates based on keyword similarity. We also removed any geographic, occupation, or product-dependent apps. To add more variety, we lastly repeated this process with three additional apps (Down Dog, Headspace, and pray.com).

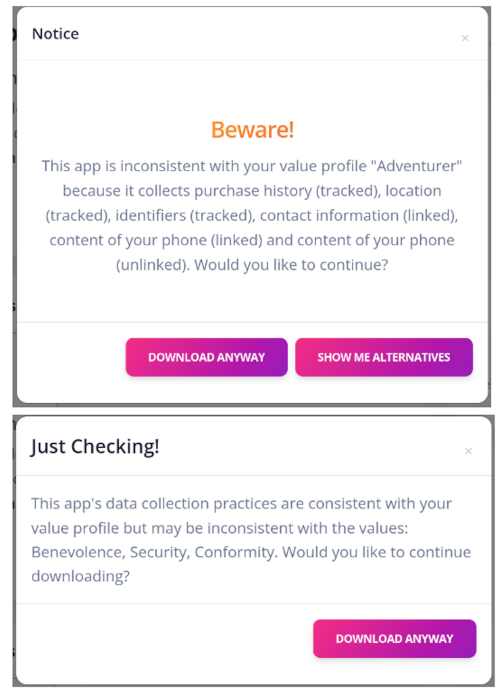

Figure 2: Examples of selective (top) and exploratory (bottom) notices.

This process resulted in 97 health and fitness apps sorted into 9 families.

The MAS and integrated VcPA were developed using JavaScript and Python's Flask framework. Recall that the VcPA involves three primary features: selective notices based on a value profile; suggesting alternative applications that are more consistent with one's selected profile; and exploratory notices (Table 1). To accomplish this in the MAS, we firstly calculated a minimal acceptability coefficient using survey data in Phase I. The coefficient is the percentage of participants accepting the data practice required by the app that the least number of survey participants in that profile would be willing to accept. Selective notices were triggered when an app's practices did not match the profile, determined as a cutoff point of <0.1 to minimize notice fatigue (Figure 2). We also added a "traffic light" system to the apps, where apps that were below 0.1 were "red" (and would trigger a notice); "yellow" if between 0.1 and 0.5; and "green" if above 0.5. Selective notices also included a button pointing to the "suggest alternatives" page, which included apps that matched the participants' profile (coefficient>0.1) in the app's family. The VcPA exploratory notices, utilized to check that the user's profile was still the best match (Figure 2), were also integrated into the Mock App Store. These notices were triggered between 3.5 minutes and 4 minutes into MAS engagement when trying to download an app.

Participants were recruited using snowball sampling over the researchers' networks in a manner like the survey. Before starting the exercise, participants were asked to complete an entry survey asking about demographic details, basic privacy attitudes, and values. Questions asking participants to rank their level of overall privacy concern and smartphone privacy concern were also included on the entry survey and an exit survey. When opening the MAS, participants were first shown the VcPA profiles along with a brief description and the top values associated with that profile (Figure 1). Participants were then requested to choose a profile based on their personal values and asked to browse the MAS. App icons and detailed descriptions were presented to enable participants to select relevant apps to download to their "virtual smartphone." They were also given the option to remove apps. Each interaction with the MAS, such as which profiles were selected, which apps were downloaded or removed, and interactions with selective and exploratory notices, was recorded. If a participant ignored a selective notice, they were asked to specify a reason, which was also recorded. Where applicable, interaction differences between the three profiles were analyzed using t-tests. To calculate the number of apps that users "download" consistent with their value profiles, we defined "consistent" apps using the minimal acceptability coefficient of >0.1. Due to unequal sample sizes, data from the entry and exit survey were analyzed using a two-sample unequal t-test.

**Phase III: Semi-Structured Interviews**
Phase III involved follow-up semi-structured interviews with a select number of Mock App Store participants to further assess the VcPA and elicit feedback from participants. At 18 interviews it was apparent that we had reached saturation (no new content) and had sufficient informational power (sufficient richness of data) (Braun and Clarke 2022; Glaser and Strauss 1999; Malterud, Siersma, and Guassora 2016; Sandelowski 1995). The interviews were 20 minutes to an hour in length, conducted in English over Microsoft Teams. They were semi-structured to allow participants sufficient space to explore their personal privacy and app decision-making process, while also covering defined issues of interest. Otter.AI (https://otter.ai) was used to generate initial transcripts, which were checked against recordings for accuracy and any identifying information. Interviews were transcribed clean verbatim, with filler words removed unless deemed necessary for understanding. Analysis was done using a process of Reflexive Thematic Analysis (TA) to identify feedback (Braun and Clarke 2022). The reflexive TA process was completed twice in NVivo, with interview transcripts shuffled for the second pass to ensure an evenly coded dataset. Theme development occurred iteratively, both during the initial pass with finalization after coding.

## Results and Discussion

Here, we report and discuss our empirical results assessing the usability and effectiveness of a VcPA. Participant interactions with VcPA notices and features, as well as participant feedback, were analyzed to established proof-of-concept for a value-centered privacy assistant based on user values and privacy preferences. We also identified areas of improvement for the VcPA, laying the groundwork for future VcPA research.

### Demographics
For the survey (Phase I), 273 participants were recruited, with 147 assigned to the Lose it! version of the survey and 126 to the OpenLitterMap version. Participants were mostly adults (ages 25-64; 204 participants); had at least a master's degree (214); identified as women (163); and were of European nationalities (176). Members of each demographic group were spread evenly between the Lose It! and OpenLitterMap versions of the survey. Following the survey, a second group of 120 participants were recruited to partake in the Mock App Store Study (Phase II). Of the participants recruited, 111 participants completed the entry survey with demographic details. Participants were primarily adults (82 participants); identified as women (63); had at least a bachelor's degree (100) and were of European nationality (67). After excluding logs from the MAS that did not include any downloaded apps, we had 77 usable participant logs. 66 participants completed the exit survey, with an additional 18 participants recruited for the semi-structured interviews. These participants were primarily adults (17 participants); of European nationality (9 participants); and all had at least a bachelor's degree. Gender was evenly split between men and women. More demographic information for all phases is available in Supplementary Table III.

**The VcPA and Value-Centered App Choices**
Overall, our results suggest that a VcPA helps participants download apps more consistently with their values. 35 of participants had a high percentage (>90%) of apps downloaded at the end of the Mock App Store exercise matching their profile, with only 11 having a low percentage (<10% match). This suggests that the VcPA's selective notices were reasonably successful at helping participants to act according to their values when making app choices. Interestingly, those who selected the Helpful Neighbor profile tended to download more apps that were consistent with their profile than the other two profiles (p=0.0003 and 0.0008). Taken together, these results support proof-of-concept for a VcPA, demonstrating how a value-centered approach to privacy could help users act in better accordance with their values.

**VcPA Profiles**
Our results concerning VcPA profiles further support using values as a basis for privacy recommendations. Our method of profile design was also mostly supported. Firstly, we found that participants leaned towards finding values associated with each profile and finding a profile that reflected them, with most participants ranking profiles as satisfactory in the exist survey. In the interviews, three participants mentioned that they felt they found a profile that was a good match for them (P09: "I suppose really, like Goal Setter was very, very, very close to perfect, probably perfect"). Secondly, profiles were also consistently interpreted. Goal Setter was (positively and negatively) seen as primarily focused on *work* goals (5 participants; P14: "Yeah, I'm more of a Not Goal Setter!") rather than goals more broadly (3 participants). In this vein, some participants saw Goal Setter as being about the destination rather than the journey or about living in the future, rather than the moment (P08: "I find myself a bit more out on the limb most of the time, and […] [that's] where I enjoy being."). Adventurer was primarily associated with individual autonomy and freedom to live one's life (4 participants), as well as being open to new experiences (3 participants), or "just go[ing] with it" (P12). Helpful Neighbor was viewed as being associated with being conforming, humble, and caring for others first and foremost.

While the profiles were well-received and consistently understood, we also identified areas of improvement. To start, while participants reported finding a profile that matched them, we also learned that many interview participants saw themselves between profiles (4 were between Helpful Neighbor and Goal Setter; 2 between Adventurer and Helpful Neighbor; 1 between Adventurer and Goal Setter; and 4 between every profile). Three participants described the profiles as "high-level" (P02) and vague, with the line between profiles not clear, and one felt that their real-life profile would be a mix of profiles. These results suggest that designing more personalized profiles could further help users act according to their values, moving us from proof-of-concept to a more optimized VcPA that could be implemented in app stores. Some participant-proposed improvements included customizable profiles to allow users to create the profile that best reflects them; a survey to sort a user into a profile; and adding more profiles. To avoid cognitive overload with an excessive number of profiles, we think that designing customizable profiles ("mix-and-match") and/or using a survey or decision tree to initially sort users into profiles would be the most promising avenues for future work. In addition, the value-privacy relationship, established by the survey data and forming the basis of the VcPA profiles, does not appear to be capturing all relevant values. The relevance of other values was present in the data in a few ways, including four interview participants who felt their relevant values would be heavily dependent on context and MAS participants describing app *usefulness*, *function*, *engagement, efficiency* and other values as critical when choosing apps in the app store. Taken together, this suggests that the ten broad Schwartz values may not be sufficient to capture *all* relevant values that may be related to privacy-decision making in a context-dependent manner. It would therefore behoove future researchers to explore VcPA profile design using methods that can capture the context-specific values (e.g., Liscio, (2022)) and assess their efficacy against those derived from Schwarz.

**Selective Notices and "Suggest Alternatives" Feature**
Selective notices and the "suggest alternatives" feature were well-received by participants, showing further proof-of-concept for the value-centered approach to privacy – in particular, that offering a more value-consistent alternative course of action can help users act in better accordance with their values. In support of this, the "suggest alternatives" feature was the most popular feature, with 69% of Mock App Store participants rating it 4 or higher (average=3.79, SD=1.28, median=4). A few interview participants expressed satisfaction with the "suggest alternatives" page, describing it as "really handy" to consider "apps with the same functionality" (P14), something they "would never think to do […] on [their] own" (P07). Most aspects (timing, content, frequency, overall) of the VcPA notices were also moderately positively received. For the notices, 80% of participants rated their overall satisfaction with VcPA notices above an average score of 3 out of 5 (average =3.64; SD=1.15; me-

dian=4). 62% rated the frequency of notices above 3 (average=3.23, SD=0.84, median=3). 89% rated the timing of the notices above 3 (average=3.98, SD=1.06, median=4).

While the "suggest alternatives" feature appeared to help participants find alternative apps, we also identified where it could be improved in future work. When asked what other features would help them select a smartphone application, some MAS participants recommended improving the ability to compare apps on the "suggest alternatives" page with added information about the app (e.g., app reviews) and app functionality. Some participants also did not think the alternatives recommended were close enough in terms of their function. Feedback gathered from the interviews echoed that in the written feedback, including difficulty comparing alternatives and a lack of function match between supposed similar apps. The "suggest alternatives" feature would therefore benefit from more app information; more ease in comparing similar apps; and better matching the similar app functions to the original app. The first limitation concerning app information was more a limitation of the MAS interface, something that would likely not be an issue if a VcPA was used in an authentic app store setting. However, a comparison feature on the selective notice itself could further enable users to make app choices consistent with their values. Matching based on descriptions, rather than keywords, could also improve the function match. In addition, while a generally well-rated feature, use of the "suggest alternatives" page was split between two extremes, with roughly a fourth participants having high engagement (using the "suggest alternatives" button >90% of the time they received a selective notice) and a fourth of participants having low engagement (clicking less than <10% of the time). This suggests that the "see alternatives" button on the selective notice was especially engaging for some participants but equally disengaging (although desirable) for others. To see how notice engagement could be improved, we explored why notices were ignored by some participants, and found that some participants were confused about the type of data being collected ("I am not sure what tracked, linked, unlinked means, so I am not sure what they collect exactly"). This was echoed in the interviews and written exit survey feedback. These results suggest that the Apple Privacy Label ontology we utilized to represent privacy preferences is difficult for users to understand, supporting other reports that the labels are not as effective as once hoped (Zhang et al. 2022; Kollnig et al. 2022). Future work using other privacy ontologies will therefore be needed to improve selective notice understandability and transparency. We could perhaps take inspiration from the "Privacy Facts" display from Kelley (2013), where we could visually display the data being collected alongside the associated values to increase user comprehension.

**Ethical Considerations and Study Limitations**

Using an established methodology from cross-cultural psychology, the TBHV, allowed us to measure values efficiently to craft VcPA profiles. However, we should be wary of reductionism by assuming that the rich and diverse landscape of human values can be fully captured in ten values, or the profiles derived from them. Evolutionary psychology forms the basis for the TBHV (Schwartz 1992), and applications of evolutionary theory to social phenomena have been heavily critiqued for explaining human behavior based on biological principles alone (Dupré 2001; Caporael and Brewer 1991). Psychology itself has been critiqued for its WEIRD (Western, Educated, Industrial, and Democratic) biases (Henrich, Heine, and Norenzayan 2010), which are relevant here for an investigation of values. Our experiment included predominantly "WEIRD" participants, with the vast majority holding a higher degree and from the Western world (Supplementary Table III). While this study aimed to establish proof-of-concept through initial prototyping and testing, VcPA profiles must be properly contextualized if we wish to have a VcPA that is helpful to all user populations. Qualitative investigations, such as interviews, and engagement with diverse voices are required to gain a richer understanding of the values involved in data privacy decisions and to develop a VcPA that is inclusive, beneficial, and ethical.

## Conclusion

This study demonstrates proof-of-concept that a VcPA system can help users make more value-centered privacy decisions. To accomplish this goal, we aimed to translate existing theoretical and conceptual VcPA work into a prototype system for choosing apps and aimed to evaluate the VcPA in terms of usability and effectiveness. To meet these aims, we conducted a mix-methods study that 1.) captured the relationship between values and privacy preferences to generate VcPA profiles; 2.) empirically measured the usability of the VcPA; and 3.) elicited user feedback. We saw that the VcPA did help participants download more apps consistent with their values. We found that a particularly well-received feature of the VcPA was "suggest alternatives," which some participants felt helped them find alternative apps of similar function that were more consistent with their values. In addition, most participants found VcPA profiles clear and could find one that they felt reflected them. Taken together, these results demonstrate proof-of-concept that a VcPA, designed based on user values rather than on privacy preferences alone, does help users make privacy decisions according to their values.

We also succeeded in identifying areas where the VcPA could be improved, thereby laying the foundation for future research into VcPAs. Future research could explore how VcPA profiles could be more tailored to each participant, perhaps by making them customizable and/or using a survey to sort users into profiles, as recommended by participants. Profiles could also be constructed using alternative context-specific theories of values. Future research could investigate designing a more streamlined "suggest alternatives" feature on the notice itself with better function match, as we found was desired by participants. Clearer presentation of values and data collection practices on selective notices, perhaps by visually displaying the data being collected alongside the associated values, could be explored to increase user comprehension. These areas of future work, in conjunction with diverse user engagement, will help bring about a VcPA that can be implemented on app stores, supporting users to make more meaningful, autonomous, and value-centered privacy decisions.

## Acknowledgements

This study described in this work was completed during Sarah's PhD, which was funded by the Science Foundation Ireland Centre for Research Training in Digitally-Enhanced Reality (d-real) under Grant No. 18/CRT/6224. The authors would like to firstly thank all the participants, without whom these studies would not have been possible and the Knowledge in AI group at VU Amsterdam for hosting Sarah.

## Declarations

### Funding

This work was funded by the Science Foundation Ireland Centre for Research Training in Digitally-Enhanced Reality (d-real) under Grant No. 18/CRT/6224.

### Competing Interests

Authors have no financial or other competing interests to report.

### Ethics

Ethics approval for this study was granted by the Ethics Committee of University of Galway on (24.09.2021/No.2021.09.003). An amendment was also approved by the Ethics Committee (15.03.2022/No.2021.09.003; Amend2203).

### Consent to Participate

Informed consent was obtained from all individual participants included in the study. If participants participated in multiple study phases, consent was re-obtained before each phase.

### Consent to Publish

All participants consented to having their demographic information reported (in aggregate) and results (including interview quotes) published either in aggregate or with an anonymous identifier.

### Data, Materials, and Code Availability

Raw data from this study is only available to the primary research team in accordance with the GDPR, the study ethics approval granted, and the content of the informed consent form. Data is available in aggregate upon request (excluding interview transcripts to protect participant anonymity).

## References

Alashoor, T.; Keil, M.; Liu, L.A.; and Smith, J. 2015. How Values Shape Concerns about Privacy for Self and Others. In Proceedings of the International Conference on Information Systems.

Braun, V. and Clarke, V. 2022. *Thematic Analysis: A Practical Guide*. Edited by A. Maher. 1st ed. London: SAGE Publications.

Caporael, L. R. and Brewer, M.B. 1991. The Quest for Human Nature: Social and Scientific Issues in Evolutionary Psychology. *Journal of Social Issues* 47(3): 1–9. https://doi.org/https://doi.org/10.1111/j.1540-4560.1991.tb01819.x.

Carter, S. E. 2021. Is Downloading This App Consistent with My Values?: Conceptualizing a Value-Centered Privacy Assistant. In Lecture Notes in Computer Science, edited by Dennehy, D.; Griva, A.; Pouloudi, N.; Dwivedi, Y.; Pappas, I.; and Mäntymäki, M. 12896 LNCS:285–91. Galway: Springer International Publishing. https://doi.org/10.1007/978-3-030-85447-8_25.

Carter, S. E. 2022. A Value-Centered Exploration of Data Privacy and Personalized Privacy Assistants. *Digital Society* 1(27). https://doi.org/10.1007/s44206-022-00028-w.

Carter, S. E.; Tiddi, I.; and Spagnuelo, D. 2022. A "Mock App Store" Interface for Virtual Privacy Assistants. In Proceedings of the First International Conference on Hybrid Human-Artificial Intelligence (HHAI2022). Amsterdam: iOS Press.

Carter, S.E. 2023. A value-centered approach to data privacy decisions. PhD dissertation, Department of Philosophy, University of Galway, IE.

Cox, A. L.; Gould, S.; Cecchinato, M.E.; Iacovides, I.; and Renfree, I. 2016. Design Frictions for Mindful Interactions: The Case for Microboundaries. In Proceed-ings of the Conference on Human Factors in Computing Systems (CHI), 1389–97. San Jose: ACM. https://doi.org/10.1145/2851581.2892410.

Das, A.; Degeling, M.; Smullen, D.; and Sadeh, N. 2018. Personalized Privacy Assistants for the Internet of Things: Providing Users with Notice and Choice. *IEEE Pervasive Computing* 17 (3): 35–46. https://doi.org/10.1109/MPRV.2018.03367733.

Dupré, J. 2001. *Human Nature and the Limits of Science*. Oxford University Press. https://doi.org/https://doi.org/10.1093/0199248060.001.0001.

Glaser, B. G. and Strauss, A.L. 1999. *The Discovery of Grounded Theory*. 1st ed. Routledge. https://doi.org/10.4324/9780203793206.

Gray, C. M.; Kou, Y.; Battles, B.; Hoggatt, J.; and Toombs, A.L. 2018. The Dark (Patterns) Side of UX Design. In Proceedings of the Conference on Human Factors in Computing Systems (CHI). 1–14. Montreal: ACM. https://doi.org/10.1145/3173574.3174108.

Henrich, J.; Heine, S.J.; and Norenzayan, A. 2010. The Weirdest People in the World? *Behavioral and Brain Sciences* 33(2–3): 61–83. https://doi.org/10.1017/S0140525X0999152X.

Kahneman, D. 2011. *Thinking, Fast and Slow*. Penguin Books.

Kelley, P.G; Cranor, L.F.; and Sadeh, N. 2013. Privacy as Part of the App Decision-Making Process. In Proceedings of the Conference on Human Factors in Computing Systems (CHI). Paris: ACM. https://doi.org/10.1145/2470654.2466466.

Killmister, S. 2017. *Taking the Measure of Autonomy: A Four-Dimensional Theory of Self-Governance*. 1st ed. New York: Routledge. https://doi.org/https://doi.org/10.4324/9781315204932.

Kollnig, K.; Shuba, A.; Van Kleek, M.; Binns, R.; and Shadbolt, N. 2022. Goodbye Tracking? Impact of IOS App Tracking Transparency and Privacy Labels. In Proceedings of the Conference on Fairness, Accountability, and Transparency (FAccT). Seoul: ACM. https://doi.org/10.1145/3531146.3533116.

Kruskal, W H., and Wallis, A.W. 1952. Use of Ranks in One-Criterion Variance Analysis. *Journal of the American Statistical Association* 47(260): 583–621. https://doi.org/10.1080/01621459.1952.10483441.

Lindeman, M. and Verkasalo, M. 2010. Measuring Values with the Short Schwartz's Value Survey. *Journal of Personality Assessment* 85(2): 170–78. https://doi.org/10.1207/s15327752jpa8502.

Liscio, E.; van der Meer, M.; Siebert, L.C; Jonker, C. M.; and Murukannaiah, P.K. 2022. What Values Should an Agent Align with?: An Empirical Comparison of General and Context-Specific Values. *Autonomous Agents and Multi-Agent Systems*. 36(23). .https://doi.org/10.1007/s10458-022-09550-0.

Liu, B.; Andersen, M.S.; Schaub, F.; Almuhimedi, H.; Zhang, S.; Sadeh, N.; Acquisti, A.; and Agarwal, Y. 2016. Follow My Recommendations: A Personalized Privacy Assistant for Mobile App Permissions. In Proceedings of the Twelfth Symposium on Usable Privacy and Security (SOUPS). Denver: USENIX.

Liu, B.; Lin, J.; and Sadeh, N. 2014. Reconciling Mobile App Privacy and Usability on Smartphones: Could User Privacy Profiles Help? In Proceedings of the International Conference on World Wide Web. Seoul: ACM. https://doi.org/10.1145/2566486.2568035.

Malterud, K.; Siersma, V.D.; and Guassora, A.D. 2016. Sample Size in Qualitative Interview Studies: Guided by Information Power. *Qualitative Health Research* 26(13): 1753–60. https://doi.org/10.1177/1049732315617444.

Mathur, A.; Acar, G.; Friedman, M.J.; Lucherini, E.; Mayer, J.; Chetty, M.; and Narayanan, A. 2019. Dark Patterns at Scale: Findings from a Crawl of 11K Shopping Websites. In Proceedings of the ACM Conference on Computer Supported Cooperative Work (CSCW). 3:81:1-81:32. Austin: ACM. https://doi.org/10.1145/3359183.

Milligan, G. W. and Cooper, M.C. 1988. A Study of Standardization of Variables in Cluster Analysis. *Journal of Classification* 5(2): 181–204. https://doi.org/10.1007/BF01897163.

Nurwidyantoro, A.; Shahin, M.; Chaudron, M.R.V.; Hussain, W.; Shams, R.; Perera, H.; Oliver, G.; and Whittle, J. 2022. Human Values in Software Development Artefacts: A Case Study on Issue Discussions in Three Android Applications. *Information and Software Technology* 141(106731). https://doi.org/10.1016/j.infsof.2021.106731.

Obie, H. O.; Hussain, W.; Xia, X.; Grundy, J.; Li, L.; Turhan, B.; Whittle, J.; and Shahin, M. 2021. A First Look at Human Values-Violation in App Reviews. In Proceeding of the International Conference on Software Engineering (ICSE), 29–38. Madrid: IEEE/ACM. https://doi.org/10.1109/icse-seis52602.2021.00012.

Perera, H.; Hussain, W.; Mougouei, D.; Shams, R.A.; Nurwidyantoro, A.; and Whittle, J. 2019. Towards Integrating Human Values into Software: Mapping Principles and Rights of GDPR to Values. In Proceedings of the International Conference on Requirements Engineering, 404–9. IEEE. https://doi.org/10.1109/RE.2019.00053.

Peters, D.; Calvo, R. A.; and Ryan. R. M. 2018. Designing for Motivation, Engagement and Wellbeing in Digital Experience. *Frontiers in Psychology* 9. https://doi.org/10.3389/fpsyg.2018.00797.

Rosson, M. B. and Carroll, J. 2002. Scenario-Based Design. In *The Human-Computer Interaction Handbook: Fundamentals, Evolving Technologies and Emerging Applications*, edited by J. Jacko and A. Sears, 1st ed., 1032–50. Lawrence Erlbaum Associates. https://doi.org/10.2307/798660.

Sandelowski, M. 1995. Sample Size in Qualitative Research. *Research in Nursing & Health* 18(2): 179–83. https://doi.org/https://doi.org/10.1002/nur.4770180211.

Sandhaus, H. 2023. Promoting Bright Patterns. In CHI '23: Designing Technology and Policy Simultaneously Workshop, 1–9. https://brightpatterns.org/.

Schaub, F.; Balebako, R.; Durity, A.L.; and Cranor, L.F. 2015. A Design Space for Effective Privacy Notices. In Proceedings of the Symposium on Usable Privacy and Security (SOUPS), 1–17. Ottawa: USEBIX. https://doi.org/10.1017/9781316831960.021.

Schwartz, S. H. 1992. Universals in the Content and Structure of Values: Theoretical Advances and Empirical Tests in 20 Countries. *Advances in Experimental Social Psychology* 25(C): 1–65. https://doi.org/10.1016/S0065-2601(08)60281-6.

Schwartz, S. H. 1994. Are There Universal Aspects in the Structure and Contents of Human Values? *Journal of Social Issues* 50(4): 19–45. https://doi.org/10.1111/j.1540-4560.1994.tb01196.x.

Schwartz, S. H. 2012. An Overview of the Schwartz Theory of Basic Values. *Online Readings in Psychology and Culture* 2(1): 1–20. https://doi.org/10.9707/2307-0919.1116.

Schwartz, S. H.; Cieciuch, J.; Vecchione, M.; Davidov, E.; Fischer, R.; Beierlein, C.; Ramos, A. et al. 2012. Refining the Theory of Basic Individual Values. *Journal of Personality and Social Psychology* 103(4): 663–88. https://doi.org/10.1037/a0029393.

Shams, R. A.; Shahin, M.; Oliver, G; Perera, H.; Whittle, J.; Nurwidyantoro, A.; and Hussain, W. 2023. Investigating End-Users' Values in Agriculture Mobile Applications Development: An Empirical Study on Bangladeshi Female Farmers. *Journal of Systems and Software* 200(111648). https://doi.org/10.1016/j.jss.2023.111648.

Solove, D. J. 2021. The Myth of the Privacy Paradox. *George Washington Law Review* 89(1): 1–51. https://doi.org/10.2139/ssrn.3536265.

Terpstra, A., Schouten, A.P.; de Rooij, A.; and Leenes, R.E. 2019. Improving Privacy Choice through Design: How Designing for Reflection Could Support Privacy Self-Management. *First Monday* 24(6). https://doi.org/https://doi.org/10.5210/fm.v24i7.9358.

Troullinou, P. 2017. Exploring the Subjective Experience of Everyday Surveillance: The Case of Smartphone Devices as Means of Facilitating" Seductive" Surveillance. PhD dissertation, Department of Computer Science, The Open University, UK.

Utz, C.; Degeling, M.; Fahl, S.; Schaub, F.; and Holz, T. 2019. (Un)Informed Consent: Studying GDPR Consent Notices in the Field. In Proceedings of the Conference on Computer and Communications Security, 973–90. London: ACM. https://doi.org/10.1145/3319535.3354212.

Warberg, L.; Acquisti, A.; and Sicker, D. 2019. Can Privacy Nudges Be Tailored to Individuals' Decision Making and Personality Traits? In Workshop on Privacy and Electronic Society (WPES), 175–97. London: ACM. https://doi.org/10.1145/3338498.3358656.

De Wet, J.; Wetzelhütter, D.; and Bacher, J. 2019. Revisiting the Trans-Situationality of Values in Schwartz's Portrait Values Questionnaire. *Quality and Quantity* 52(2): 685–711. https://doi.org/10.1007/s11135-018-0784-8.

Zhang, S.; Feng, Y.; Yao, Y.; Cranor, L. F.; and Sadeh, N. 2022. How Usable Are IOS App Privacy Labels? In Proceedings on Privacy Enhancing Technologies, 204–28. https://doi.org/10.56553/popets-2022-0106.